\documentclass[preprint,aps]{revtex4}
\usepackage{graphicx,amsfonts}
\usepackage{dcolumn}
\usepackage{bm}

\begin{document}
\title {\Large Spontaneous Symmetry Breaking in Noncommutative Field Theory}
\author{H. O. Girotti}
\affiliation{Instituto de F\'{\i}sica, Universidade Federal do Rio Grande
do Sul, Caixa Postal 15051, 91501-970 - Porto Alegre, RS, Brazil}
\email{hgirotti@if.ufrgs.br}
\author{M. Gomes}
\author{A. Yu. Petrov}
 \altaffiliation[Also at]{ Department of Theoretical Physics,
Tomsk State Pedagogical University
Tomsk 634041, Russia
(email: petrov@tspu.edu.ru)}
\author{V. O. Rivelles}
\altaffiliation[]{ Center for Theoretical Physics,
Massachusetts Institute of Technology,
Cambridge, MA 02139-4307, USA
(email: rivelles@lns.mit.edu)}
\author{A. J. da Silva}
\affiliation{Instituto de F\'{\i}sica, Universidade de S\~{a}o Paulo,
 Caixa Postal 66318, 05315-970, S\~{a}o Paulo - SP, Brazil}
\email{mgomes, petrov, rivelles, ajsilva@fma.if.usp.br}

\begin{abstract}
The spontaneous symmetry breaking of rotational $O(N)$ symmetry in
noncommutative field theory is investigated in a 2+1 dimensional model
of scalar fields coupled through a combination of quartic and
sextuple self-interactions. There are five possible orderings of the 
fields in the sextuple interaction and two for the quartic
interaction.  At one loop, we prove that for some choices of these
orderings there is the absence of IR/UV mixing and the appearance of
massless excitations. A supersymmetric extension of the model is also
studied.  Supersymmetry puts additional constraints on the couplings
but for any given $N$ there is a Moyal ordering of the superfields
for which the requirement for the existence of Goldstone bosons is
satisfied. For some ordering and when  $N\rightarrow\infty$ we find
evidence that the model is  renormalizable to all orders in perturbation
theory. We also consider a generic chiral model in $3+1$ dimensions whose 
superpotential is
invariant under local gauge transformations. We find that for any
value of $N$ there is no one loop correction to the pion mass and
that, at two loops, there are no pion mass corrections for slowly
varying superfields so that Goldstone theorem holds true. We also find
a new purely noncommutative coupling which gives contributions
starting at order $N-2$ loops.  
\end{abstract}

\maketitle
\newpage

\section{INTRODUCTION}

Noncommutative quantum field theory has been intensively studied
during the last years (for a review see \cite{review}).  At 
first, the idea of noncommutativity was motivated by the hope that its
introduction (and the ensuing nonlocality) would allow for the
construction of theories with improved ultraviolet behavior.  This
expectation was not fulfilled as noncommutative theories exhibit
ultraviolet divergences of the same sort as the ones found in the
commutative situation.  However, nowadays there are other motivations,
coming mainly from string theory and quantum gravity, which keep a
very keen interest on the subject. Being intrinsically nonlocal these 
models present many unusual aspects which deserve painstaking
investigation. Its main characteristic is the appearence of an
infrared (IR) singularity (also  referred to as UV/IR mixing\cite{Minw})
which may ruin some of the properties that a well defined quantum field 
theory is required to possess. For instance,
perturbative renormalizability is usually lost in noncommutative
theories although it is regained in the supersymmetric case
(see\cite{NCWZ,sigma} and \cite{Toumbas,Ruiz1} for studies in non-gauge and
gauge theories, respectively).  

Another aspect which deserves a better understanding is the
effect of these IR singularities  on the mechanism of 
symmetry breakdown and the appearance of Goldstone bosons in
noncommutative field theory.  Previous studies unveiled some
interesting features. For the four dimensional linear sigma model it
has been shown that, at one loop, spontaneous breakdown may occur for
the $U(N)$ theory but not for $O(N)$ if $N  \not= 2$. It is also worth
mentioning that the Goldstone theorem holds only if the field 
ordering in the quartic Moyal product is consistent with local
symmetry \cite{KK}. Other properties of the
$O(2)$ case have also been studied in \cite{Petriello,Ruiz} and the results for
the $U(N)$ case have been extended to two loops \cite{Liao}. Attempts
to go to higher loops requires an IR regulator 
which can not longer be removed\cite{Sarkar}.  Thus, only certain
noncommutative extensions preserve the main features of their commutative
counterpart. One may ask whether this depends upon the dimensionality of
space-time or on the nature of the interaction or on
both. This paper provides further understanding on this problem
by means of some specific examples.

 We shall start by studying the spontaneous symmetry breaking of the
 continuous $O(N)$ symmetry in a three dimensional theory and in a 
supersymmetric version of it, with the aim learning about the role played
by supersymmetry. 

In the nonsupersymmetric case
we find that there is class of field orderings for which no UV/IR mixing
occurs at one loop. Unlike the four dimensional case it does not
require gauge invariant couplings. Interesting enough, the condition for
the elimination of UV divergences, in the planar sector of the pion two
point function, also secures the cancellation of the would be IR
divergence, in the nonplanar sector, and, at the same time, enforces
the appearance of massless excitations.  
This mechanism is absent in the four dimensional case. We remark that it is
possible to extend the nonsupersymmetric model by adding a purely 
noncommutative interaction, i.e., one with no
commutative analog. It gives loop contributions starting at order
$N-2$ so it is not relevant for our one loop results. It also exists in
superspace but gives no contribution to the supersymmetric
case. Its role outside the context of spontaneous symmetry breaking
deserves further investigation. 

While in four dimensions supersymmetry removes a dangerous IR
quadratic singularity, in our 
three dimensional case it just restricts further the class of allowed
models. Moreover, for a specific Moyal ordering we find evidence of
its renormalizability to any order in the limit $N\rightarrow \infty$. 

As far as noncommutative four dimensional theories are concerned, we shall
be dealing with effective supersymmetric
field models which arise as the low energy limit of compactified
string or M theory. Usually the tree level exchange of heavy fields
leads to nonrenormalizable interactions in the effective potential of
the light fields. In the supersymmetric case, the decoupling of the
heavy fields leads to corrections to the effective superpotential
\cite{Ross} and to the K\"ahler potential of the light fields
\cite{Cvetic}.  On the other hand, the dynamics of D3-branes may be
described, for slowly varying fields, by a Born-Infeld type 
action in which the transversal radial excitations are a set of scalar
fields \cite{Tseytlin}. In the supersymmetric case one finds a chiral
superfield in a specific superpotential 
\cite{Kuzenko}. In view of this it is natural to analyze the
noncommutative versions of such theories and to study the
validity of the 
Goldstone theorem for them, since they provide extensions of realistic
string models.  We assume that the interaction can be approximated
by a general superpotential \cite{petrov} which allows for a local
$O(N)$ gauge symmetry. It will be shown that, at one loop, 
there is no corrections to the pion mass so that Goldstone theorem
holds true. Then, using supersymmetry alone, we go to two 
loops and show that for slowing varying superfields there are, again,
no pion mass corrections. Thus, supersymmetry enables us to
transpose the $N=2$ barrier found for the purely bosonic quartic
interaction\cite{KK}.

This paper is organized as follows. In Section II the noncommutative
version of the $O(N)$  scalar model with quartic and sextuple
interactions is presented and the spontaneous symmetry breaking is
studied. Section III contains a discussion of its supersymmetric
version in superspace. Finally, in Section IV we discuss the
spontaneous symmetry breaking of a chiral superfield in a generic 
superpotential in four dimensions. The conclusions are left to
Section V.

\section{The noncommutative $\phi^6$ model}

We start our analysis by considering the possible spontaneous
breakdown of the $O(N)$ symmetry in a $d$ dimensional model described
by the action (subscripts from the beginning and the middle of the
latin alphabet run from 1 to $N$ and from 1 to $N-1$, respectively)

\begin{eqnarray}
\label{action1}
S&=&\int d^d x \Big[-\frac{1}{2}\phi_a\Box\phi_a+\frac{\mu^2}{2}\phi_a\phi_a
\nonumber\\&-&
\frac{g}4 \Big ( l_1 \phi_a*\phi_a*\phi_b*\phi_b
+l_2\phi_a*\phi_b*\phi_a*\phi_b\Big)
\nonumber\\
&-&
\frac{\lambda}{6}\Big(h_1\phi_a*\phi_a*\phi_b*\phi_b*\phi_c*\phi_c+
h_2\phi_a*\phi_a*\phi_b*\phi_c*\phi_c*\phi_b+\nonumber\\&+&
h_3\phi_a*\phi_a*\phi_b*\phi_c*\phi_b*\phi_c+h_4\phi_a*\phi_b*\phi_c*\phi_a*\phi_b*\phi_c+
\nonumber\\&+&
h_5\phi_a*\phi_b*\phi_c*\phi_a*\phi_c*\phi_b
\Big)
\Big].
\end{eqnarray}

\noindent
The $\ast$ indicates the Moyal product which satisfies

\begin{eqnarray}
& &\int d^d x \phi_1(x)*\phi_2(x)\ldots \phi_n(x) =\int
\frac{d^dk_1}{(2\pi)^d}\ldots \frac{d^dk_n}{(2\pi)^d}(2\pi)^d\delta^d(k_1+\ldots+k_n)
\times\nonumber\\&\times&
\exp(i\sum_{i<j}k_i\wedge k_j)\phi_1(k_1)\phi_2(k_2)\ldots \phi_n(k_n).
\end{eqnarray}

\noindent
Here, $k_i\wedge k_j=\frac{k_{i\mu}k_{j\nu}\theta^{\mu\nu}}2$, and
$\theta^{\mu\nu}=-\theta^{\nu\mu}$ is the antisymmetric constant
matrix characterizing the underlying noncommutativity.  The
$l_1,l_2,h_1,h_2\ldots h_5$ are real numbers satisfying the conditions
$l_1+l_2=1$ and $h_1+h_2+\ldots+h_5=1$,  so that there are two quartic
and five sextuple independent interaction couplings.  For constant fields, obeying
the condition $\phi_a\phi_a=a^2$, the minimal value of the action is
achieved for

\begin{equation}\label{1}
a^2= \frac{1}{2\lambda} \left ( -g + \sqrt{g^2+ 4\mu^2\lambda}\right ).
\end{equation}

 Assuming that the field
configuration which minimizes the action is of the form $(0,0,\ldots,
a)$ we redefine the fields, $(\pi_i,\sigma+a)$ so that the new fields $
\sigma,\pi_i$ have zero vacuum expectation values.  In terms of these new
fields the action takes the form

\begin{eqnarray}
\label{action4}
S&=&\int d^d x
\Big\{-\frac{1}{2}\pi_i\Box \pi_i
-\frac{1}{2}\sigma(\Box-m^2)\sigma-(2\lambda a^3+ g a)\sigma*\pi_i*\pi_i-(\frac{10}{3} \lambda a^3+g a) \sigma*\sigma*\sigma
\nonumber\\&-&
\Big[(\frac{\lambda}{6}\alpha a^2+\frac{g}4l_1)\pi_i*\pi_i*\pi_j*\pi_j+
(\frac{\lambda}{6}(3-\alpha) a^2+\frac{g}4l_2)
\pi_i*\pi_j*\pi_i*\pi_j\nonumber \\
&+&
(\frac{\lambda}{6} a^2\beta +\frac{g}2 l_1) \sigma*\sigma*\pi_i*\pi_i+(\frac{\lambda}{6} a^2(18
-\beta)+ \frac{g}2 l_2)\sigma*\pi_i*\sigma*\pi_i\nonumber\\&+&
+(\frac{5}{2} \lambda a^2
+\frac{g}4) \sigma*\sigma*\sigma*\sigma
\Big]+\ldots\Big \},\label{ac0}
\end{eqnarray}

\noindent
where $m^2= 4\mu^2-2 g a^2= 4\lambda a^4+2g a^2$, 
the dots denote terms of fifth and sixth order in the fields,
$\alpha=3h_1+2(h_2+h_3)+h_5$, 
and $\beta=18h_1+14h_2+12h_3+6h_4+8h_5$. 
Notice that condition (\ref{1}) implies that the $\pi_i$ fields (pions)
are massless  in the tree approximation, in accord with the
Goldstone theorem. 

From the action (\ref{ac0}) we can obtain the momentum space Feynman
rules. For the propagators we have

\vspace*{0.3cm}

\begin{eqnarray}
&&<\sigma(p_1)\sigma(p_2)>= (2\pi)^d \delta(p_1+p_2) \frac{i}{p_{1}^2-m^2},\\
&&<\pi_i(p_1)\pi_j(p_2)>= (2\pi)^d \delta(p_1+p_2)
\frac{i\delta_{ij}}{p_{1}^{2}},
\end{eqnarray}

\noindent
whereas the vertices carry trigonometric factors:

\begin{eqnarray}
& &\pi_i(p_1)\pi_j(p_2)\pi_k(p_3)\pi_l(p_4) 
\to -i\rho_1 [\cos(p_1\wedge p_2)\cos(p_3\wedge p_4)\delta_{ij}\delta_{kl}\nonumber\\
&&+
\cos(p_1\wedge p_3)\cos(p_2\wedge p_4)\delta_{ik}\delta_{jl}+
\cos(p_1\wedge p_4)\cos(p_2\wedge p_3)\delta_{il}\delta_{kj}]
\nonumber\\&-&
i\rho_2[\cos(p_1\wedge p_3+p_2\wedge
p_4)\delta_{ij}\delta_{kl}
+\cos(p_1\wedge p_2+p_3\wedge
p_4)]\delta_{ik}\delta_{jl}\nonumber\\& &+
\cos(p_1\wedge p_2+p_4\wedge p_3)\delta_{il}\delta_{kj}],\\
[0.3cm]
& &\pi_j(p_1)\pi_j(p_2)\sigma(p_3)\sigma(p_4) \to 
-i[\rho_3\cos(p_1\wedge p_2)\cos(p_3\wedge p_4)+\nonumber\\&+&
\rho_4\cos(p_1\wedge p_3+p_2\wedge p_4)],\\
[0.3cm]
& &\sigma(p_1)*\pi_i(p_2)*\pi_i(p_3) \to
-i(4\lambda a^3+2 g a)\cos(p_2\wedge p_3),
\end{eqnarray}

\noindent
where $\rho_1=\frac{4\lambda}{3}a^2 \alpha+2 gl_1$, $\rho_2=(3-\alpha)\frac{4\lambda}{3}a^2+ 2 g l_2$, $\rho_3=\frac{2\lambda}{3}a^2\beta+ 2 g l_1$ and $\rho_4=\frac{2\lambda}{3}a^2(18-\beta)+2gl_2$.

To study quantum corrections to the pion's mass we should introduce
the renormalizations $\phi_a\to (1+\delta_z)^{1/2}\phi_a$, $\mu^2\to
\mu^2+\delta_{\mu^2},\,\lambda\to\lambda+\delta_{\lambda}$ and $g \to
g+\delta_g$.  The reparametrizations of the ``relative'' couplings
$l$'s and $h$'s (i. e. $l_i\rightarrow l_i+\delta l_i$ and
$h_i\rightarrow h_i+\delta h_i$ with $\sum \delta
l_i=\sum\delta h_i=0$) are, of course, also done but they do not show up
in our calculations.  The corresponding counterterm Lagrangian is,
therefore,

\begin{eqnarray}
L_{ct}&=&
-\frac{1}{2} \pi_i(\delta_z\Box-\delta_{\mu^2}+\delta_g a^2+
\delta_{\lambda}a^4)\pi_i-
\frac{1}{2}\sigma(\delta_z\Box-\delta_{\mu^2}+3\delta_ga^2+ 5\delta_{\lambda}  a^4)\sigma \nonumber\\&+&
(\delta_{\mu^2} a-\delta_g a^3-\delta_{\lambda} a^5)\sigma
+ \ldots ,
\end{eqnarray}

\noindent
where the ellipsis stands for other interaction terms obtained from
 (\ref{ac0}) by replacing $\lambda$ and $g$ by $\delta_\lambda$ and $\delta_g$,
 respectively. Some of these counterterms are depicted in Fig. 1, where the $\sigma$ and $\pi$ propagators are represented by
continuous and dashed lines, respectively.

We begin our one-loop analysis of spontaneous breaking of the $O(N)$
symmetry in $2+1$ dimensions by considering the condition for the
vanishing of the vacuum expectation value of $\sigma$ (the gap
equation). It is found to read  (in the remaining of this paper all
superficially divergent integrals are implicitly assumed to be regularized. The
precise form of the regularization is irrevelant, as far as it obeys the
usual (additive) rules employed in the calculation of Feynman amplitudes) 

\begin{equation}
\delta_{\mu^2}-\delta_g a^2-\delta_{\lambda}a^4= (10 a^2\lambda + 3g)
\int \frac{d^3k}{(2\pi)^3}\frac{i}{k^2-m^2}+(N-1)(2a^2\lambda+ g)
\int\frac{d^3k}{(2\pi)^3}\frac{i}{k^2},\label{2}
\end{equation}

\noindent
thus fixing the above linear combination of counterterms. It should be
observed that this combination coincides with the $\pi$ field mass
counterterm. The gap equation (\ref{2}) is graphically represented in
Fig 2. 

We next examine the one-loop corrections to the pion's two point function
which are shown in Fig 3.  We denote the contribution from the graph
with dashed line loop as $I_1(p)$, that from the solid line loop as
$I_2(p)$, and that from the loop with two internal lines as $I_3(p)$. One has 
that

\begin{eqnarray}
I_1(p)&=&\delta^{ij}\{\frac{\lambda a^2}{3}[(2 N-4)\alpha+12] +g(Nl_1+2 l_2)\}
\int\frac{d^3k}{(2\pi)^3}\frac{1}{k^2}\nonumber\\
&+&
\delta^{ij}\{\frac{2\lambda a^2}{3}
[\alpha+(3-\alpha)(N-1)]+ gl_1+gl_2(N-1)\}\int\frac{d^3k}{(2\pi)^3}
\frac{e^{2ik
\wedge p}}{k^2},
\\[0.3cm]
I_2(p)&=&
\delta^{ij}(\frac{1}{3}\lambda a^2\beta+ g l_1)
\int\frac{d^3k}{(2\pi)^3}\frac{1}{k^2-m^2}
+\delta^{ij}[\frac{1}{3}\lambda a^2(18-\beta)+gl_2]
\int\frac{d^3k}{(2\pi)^3}\frac{e^{2ik\wedge p}}{k^2-m^2},
\\[0.3cm]
I_3(p)&=&4\delta^{ij}(2\lambda a^3+g a)^2
\int\frac{d^3k}{(2\pi)^3}\frac{\cos^2(k\wedge p)}{(k+p)^2(k^2-m^2)}.
\label{3}
\end{eqnarray}

\noindent
Therefore, at one-loop  altogether we have

\begin{eqnarray}
&&\delta^{ij}\left \{ \{\frac {2\lambda a^2}{3}[6-\alpha+ (\alpha-3)(N-1) ]
+g(Nl_1+2l_2-N+1)
\right . \}\int\frac{d^3 k}{(2\pi)^3}\frac{1}{k^2}-[10\lambda a^2\nonumber \\
&&+
(3-l_1) g
-\frac {\lambda a^2}{3}\beta]\int\frac{d^3 k}{(2\pi)^3}\frac{1}{k^2 -m^2}+\{\frac{2\lambda a^2}{3}
[\alpha+(3-\alpha)(N-1)]+ gl_1+gl_2(N-1) \}\nonumber\\
&&\int\frac{d^3 k}{(2\pi)^3}\frac{\cos(2k\wedge p)}{k^2}+ [\frac{1}{3}\lambda a^2(18-\beta)+gl_2]\int\frac{d^3 k}{(2\pi)^3}\frac{\cos(2k\wedge p)}{k^2 -m^2}+4(2\lambda a^3+g a)^2
\nonumber \\
&&\left.\int\frac{d^3k}{(2\pi)^3}\frac{\cos^2(k\wedge p)}{(k+p)^2(k^2-m^2)}+\delta_{z}
p^2\right \}.\label{3a}
\end{eqnarray}

\noindent
As $\int d^3 k \left( \frac1{k^2 - m^2}-\frac1{k^2}\right)$ is finite,
from the above results it follows that the UV divergences in the
planar contributions may be eliminated if

\begin{eqnarray}
&&\frac {2\lambda a^2}{3}[6-\alpha+ (\alpha-3)(N-1) ]+g(Nl_1+2l_2-N+1)
=10\lambda a^2+
(3-l_1) g
-\frac {\lambda a^2}{3}\beta.
\label{4}
\end{eqnarray}

\noindent
To also prevent the  IR/UV mixing in the nonplanar parts one must have

\begin{eqnarray}
&&\frac{2\lambda a^2}{3}
[\alpha+(3-\alpha)(N-1)]+ gl_1+gl_2(N-1)=-[\frac{1}{3}\lambda a^2(18-\beta)+gl_2].
\label{5}
\end{eqnarray}

Actually, we notice that equation (\ref{4}) implies (\ref{5}) and
viceversa, i. e., UV divergences and IR/UV mixing in the pions two
point functions are simultaneously eliminated.  Moreover, besides
leading to a finite result, condition (\ref{4}) (or (\ref{5})) also
secures that the resulting expression vanishes for $p=0$.  Indeed,
after imposing Eq.  (\ref{4}), the expression (\ref{3a}) may be
rewritten as

\begin{eqnarray}
&&\delta^{ij}\left \{[\frac{1}{3}\lambda a^2(18-\beta)+gl_2]\int\frac{d^3k}{(2\pi)^3}\frac{m^2\cos(2p\wedge k)}{k^2(k^2-m^2)}-[10\lambda a^2+ (3-l_1) g
-\frac {\lambda a^2}{3}\beta]\right .\nonumber \\
&&\times \int\frac{d^3k}{(2\pi)^3}\frac{m^2}{k^2(k^2-m^2)}
\left .+4(2\lambda a^3+g a)^2
\int\frac{d^3k}{(2\pi)^3}\frac{\cos^2(k\wedge p)}{(k+p)^2(k^2-m^2)}\right \},
\end{eqnarray}

\noindent
which is finite and vanishes at $p=0$. The fact
This result is a peculiarity of
the 2+1 dimensional world where there are at most linear UV
divergences (for the one loop graphs that we have considered). To see
why this is so notice that in the commutative version of the model the
sum

\begin{equation}
\mbox{pion's mass counterterm} +I_1(0) + I_2(0) \label{6}
\end{equation}

\noindent
is finite, namely, no infinite pion's wave function renormalization is
necessary. Now, it is clear that if by adjusting the parameters of the
model a subsum of the integrals occurring in (\ref{6}) is made finite
then with the same choice of parameters the sub-sum of the remaining
integrals will also be finite. In the noncommutative model that we are
dealing with the two sets of integrals correspond to the planar and
non planar parts of the graphs contributing to (\ref{6}).  The
appearance of the Goldstone bosons is a straightforward consequence of
the fact that the pion's two point function is  finite at $p=0$
(no infrared divergence). Therefore $p$ can be made zero directly on
the integrands of (\ref{3a}) leading to the same expression as in the
commutative case.

In 3+1 dimensions $\lambda=0$ is necessary for renormalizability, the
divergences are quadratic and although (\ref{3a}) still vanishes at
$p=0$ the $UV$ convergence of the planar part demands, for general
$p$, that the left hand side of (\ref{4}) be equal to

\begin{equation}
\frac{2(2 \lambda a^3+ g a)^2}{m^2}.
\end{equation}

\noindent
This new equation strongly restricts the dynamics so that class of
allowed models requires $N=2$ and $l_1=2$ as proved in \cite{KK}.

Returning to the model under analysis, we observe that for a
given $N$, equation (\ref{4}) expresses $\lambda$ in terms of $g$, for general values of $\alpha$, $\beta$ and
$l_1$. However for arbitrary non zero 
$\lambda$ and $g$ (\ref{4})  is satisfied if

\begin{equation}
l_1= \frac{N}{N-1}\qquad \mbox{and}\qquad \beta= 12+ 6 N -2\alpha(N-2).
\end{equation}

To complete the investigation on the existence of the IR/UV mixing one
has to study also the behavior of the other $n$ point functions which have
positive superficial degree of divergence. In the case of the $\sigma$
field two point function one finds that the vanishing of infrared
quadratic divergences now demands that

\begin{equation}
10\lambda a^2 +g + (N-1) [\frac{\lambda a^2}3(18-\beta)+ g l_2]= 0.
\end{equation}

The discussion may be extended to the one-loop calculations of the three
and four point functions.  One finds that the relevant contributions arise
from the vertices with five and six fields. In that situation,  the new
 parameters which come into the play are enough to eliminate the possible
IR/UV mixing. One may conclude that, as far as renormalizability is
concerned, the model is consistent up to  one-loop, for any value
  of $N$.

It is possible to generalize the action (\ref{action1}) by adding to it an
interaction which is purely noncommutative in character and whose
existence is due to the $O(N)$ symmetry. For a given $N$ it has the form 

\begin{equation}
\label{epsilon}
\int d^d x \,\,\,  \epsilon^{a_1 \dots a_N} \phi_{a_1} * \dots * \phi_{a_N}.
\end{equation}

\noindent
Clearly, (\ref{epsilon}) vanishes in the commutative case, while in
the noncommutative one it is only nonvanishing for even $N$. After
the shift, it contributes to (\ref{action4}) with  a term like 

\begin{equation}
\label{eps}
N \epsilon^{i_1 \dots i_{N-1}} \pi_{i_1} * \dots * \pi_{i_{N-1}} * 
\sigma.
\end{equation}

\noindent
For $N=4$ this contribution starts at two loops, whereas for general $N$
it starts at $N-2$ loops. So, the modification of the action implied by the addition of (\ref{epsilon}) does not alter our former conclusions, which were derived at one loop level, but may become relevant at higher loop levels. 

\section{Supersymmetric Version  in 2+1 dimensions}

A simple supersymmetric extension of the model studied in the last
section is provided by the superfield action

\begin{equation}
S=\int d^5 z \Big({\frac12} {\Phi}_a D^2 {\Phi}_a +{\frac12} \mu {\Phi}_a 
{\Phi}_a-
\frac{g}{4}
[f{\Phi}_a*{\Phi}_a*{\Phi}_b*{\Phi}_b
+(1-f){\Phi}_a*{\Phi}_b*{\Phi}_a*{\Phi}_b] \Big).
\label{n1}
\end{equation}

\noindent
Here, $D=\frac{\partial\phantom a}{\partial \theta}- i\bar \theta\not
\!\partial$, $\bar D =\gamma_o D$, $\theta_\alpha$, $\alpha=1,2$
($\bar\theta\equiv \theta \gamma^0$) are Grassmann variables,
$D^2=\frac12\bar D D$ and the superfield $\Phi$ has the decomposition

\begin{equation}
\Phi=\phi+\bar\theta \psi+\frac{\bar\theta\theta}{2}F,
\end{equation}

\noindent
where $\psi$ is a $N$ component Majorana spinor and $\phi$ and $F$ are
($N$-component) scalar fields. In terms of field components the Lagrangian reads

\begin{eqnarray}
{\cal L}&=& \frac12\partial_\mu\phi\partial^\mu\phi+ \frac{i}2\bar \psi 
\not\! \partial\psi + \frac12 F^2-\mu F\phi+\frac{\mu}2 \bar \psi\psi
+\frac{g}2f
[F_a\ast\phi_a+\phi_a\ast F_a]\ast \phi_b\ast\phi_b\nonumber\\
&\phantom a& +\frac{g}2(1-f) [F_a\ast\phi_b+\phi_a\ast F_b]\ast 
\phi_a\ast\phi_b-\frac{g}2f\bar\psi_a\ast\psi_a\ast\phi_b\ast\phi_b
\nonumber\\
&\phantom a&-  \frac{g}4f [\bar\psi_a\ast\phi_a +\phi_a\ast\bar\psi_a]
\ast[\psi_b\ast\phi_b +\phi_b\ast\psi_b]\nonumber \\
&\phantom a& -\frac{g}{2}(1-f)[ \bar\psi_a\ast\psi_b\ast\phi_a\ast\phi_b
-\psi_a\ast\bar\psi_b\ast\phi_a\ast\phi_b+ \bar\psi_a\ast\phi_b\ast\psi_a\ast\phi_b].
\end{eqnarray}

\noindent
By integrating out $F$ (or, alternatively, by eliminating it through the
use of the equations of motion) one obtains a quartic and sextuple
self-interactions with definite strengths.

In what follows we will work directly with the action
(\ref{n1}). Classically, it possesses $O(N)$ symmetry and its
potential part has a minimum at a constant value of the superfield,
$|a|=\sqrt{\frac{\mu}{g}}$.  As in the nonsupersymmetric case, to
break the $O(N)$ symmetry we suppose that one of fields (for the sake
of concreteness -- ${\Phi}_N$) has a non-zero vacuum expectation $a$.
All other fields have zero vacuum expectations. Then we make the
changes $\Phi_a=\pi_i$ for $i\neq N$ and $\Phi_N=\sigma+a$. The
superfields 
$\pi_i,\sigma$ have zero vacuum expectation values. In terms of the
new variables the action can be cast as

\begin{eqnarray}
\label{ac}
S[\pi_i,\sigma]&=&\int d^5 z \Big({\frac12} \pi_iD^2\pi_i+{\frac12}
\sigma(D^2-2\mu)\sigma\nonumber\\&-&
\frac{g}{4}\sigma*\sigma*\sigma*\sigma-g a\sigma*\sigma*\sigma -
g a\pi_i*\pi_i*\sigma-\nonumber\\&-&
\frac{g}{4}[f\pi_i*\pi_i*\pi_j*\pi_j+(1-f)\pi_i*\pi_j*\pi_i*\pi_j]-
\frac{g}{2}[f\pi_i*\pi_i*\sigma*\sigma+\nonumber\\&+&(1-f)\pi_i*\sigma*\pi_i*\sigma]
\Big).
\end{eqnarray}

\noindent
The counterterm Lagrangian is 

\begin{eqnarray}
L_{ct}&=&{\frac12} {\delta}_z\pi_i D^2\pi_i+{\frac12} 
\delta_z\sigma D^2\sigma+\frac12(-a^2\delta_{g}+{\delta}_{\mu})\pi_i\pi_i
+{\frac12}(-3a^2\delta_{g}+\delta_{\mu})\sigma^2
+(\delta_{\mu} a -\delta_{g} a^3)\sigma\nonumber\\&-&
\frac{\delta_{g}}{4}\sigma*\sigma*\sigma*\sigma-\delta_{g}\sigma*\sigma*
\sigma a-\delta_{g}\pi_i*\pi_i*\sigma a
\nonumber\\&-&
\frac{\delta_{g}}{4}[f\pi_i*\pi_i*\pi_j*\pi_j+(1-f)\pi_i*\pi_j*\pi_i*\pi_j]-
\frac{\delta_{g}}{2}[f\pi_i*\pi_i*\sigma*\sigma+\nonumber\\&+&(1-f)
\pi_i*\sigma*\pi_i*\sigma],
\end{eqnarray}

\noindent
the renormalization being done through the replacements $\pi_i\to
(1+\delta_z)\pi_i$, $\sigma\to(1+\delta_z)\sigma$,
$\mu\to\mu+\delta\mu$, $g\to g+\delta g$ and 
$f\to f+\delta f$.  The propagators
corresponding to the action (\ref{ac}) at $a=\sqrt{\mu/g}$ are

\begin{eqnarray}
&&<\pi_i(x_1, \theta_1)\pi_j(x_2, \theta_2)>=-i\frac{D^2}{\Box}\delta_{ij}\delta^5(z_1-z_2),\\
&&<\sigma(x_1, \theta_1)\sigma(x_2, \theta_2)>=-i\frac{D^2+m}{\Box+m^2}\delta_{ij}\delta^5(z_1-z_2),
\end{eqnarray}

\noindent
where  $\delta^5(z_1-z_2)=\delta^3(x_1-x_2)
\delta(\bar\theta_1 - \bar \theta_2)\delta(\theta_1-\theta_2)$
 and $m=2g a^2$. We adopt a graphical notation similar to the
one in section II. Thus, we represent the $\pi$ field propagator by a
dashed line, and the $<\sigma\sigma>$ propagator by a solid one. The
trigonometric factors associated to the cubic and quartic vertices are
the same as in the previous section and we will not list them
here. Now, however, we have to make the following identifications

\vspace*{0.5cm}

$\pi$ field bilinear counterterm: $i\delta_{ij}(\delta_{\mu}-\delta_g a^2+ \delta_z D^2)$,

\vspace*{0.5cm}

$\sigma$ field bilinear counterterm: $i(\delta_\mu - 3 \delta_g a^2+ \delta_z D^2)$,

\vspace*{0.5cm}

$\sigma$ field tadpole counterterm: $i(-\delta_g a^3+ \delta_\mu a)$.

\vspace*{0.5cm}

The vanishing of the  vacuum expectation of the $\sigma$ superfield leads
to the the gap equation

\begin{eqnarray}
3g a \int\frac{d^3 k}{(2\pi)^3}\frac{1}{k^2-m^2}+g a(N-1)
\int\frac{d^3 k}{(2\pi)^3}\frac{1}{k^2}+i(\delta_{\mu}a-\delta_{g}a^3)=0. 
\end{eqnarray}

\noindent
As before,  this condition  is not  affected  by the noncommutativity and 
implies that

\begin{eqnarray}
\delta_{\mu}a-\delta_{g}a^3=g a (N-1)
\int\frac{d^3 k}{(2\pi)^3}\frac{i}{k^2}+3g a 
\int\frac{d^3 k}{(2\pi)^3}\frac{i}{k^2-m^2}.
\end{eqnarray}

Concerning the $\pi$ superfield two point function, one finds that,
graphically, the  condition for the  cancellation of divergent corrections
to $<\pi_i\pi_j>$ propagator  is  the same as in Fig. 3.
Adopting the same convention as in the previous case, it reads

\begin{eqnarray}
I_1(p)+I_2(p)+I_3(p)+I_{ct}= \mbox{finite}.
\end{eqnarray}

\noindent
An explicit  calculation yields

\begin{eqnarray}
I_1(p)&=&\delta^{ij}g[fN+2(1-f)]\int\frac{d^3
k}{(2\pi)^3}\frac{1}{k^2}\nonumber\\&+&
\delta^{ij}g[f+(1-f)(N-1)]
\int\frac{d^3 k}{(2\pi)^3}\frac{e^{2ik\wedge p}}{k^2}, \nonumber\\
I_2(p)&=&\delta^{ij}g f
\int\frac{d^3 k}{(2\pi)^3}\frac{1}{k^2-m^2}\nonumber\\&+&
\delta^{ij}g(1-f)
\int\frac{d^3 k}{(2\pi)^3}\frac{e^{2ik\wedge p}}{k^2-m^2}, \nonumber\\
I_3(p)&=&2g^2 a^2\delta^{ij}
\int\frac{d^3k}{(2\pi)^3}
\frac{\cos^2(k\wedge p)}{k^2(k^2-m^2)}(D^2+m).
\end{eqnarray}

\noindent
Hence, the coefficient of  $D^2$ turns out to be

\begin{eqnarray}
\delta_z=2g^2 a^2
\int\frac{d^3k}{(2\pi)^3}
\frac{\cos^2(k\wedge p)}{k^2(k^2-m^2)}= 
-\frac{g^2 a^2}{\sqrt{2}\pi m}+O( \theta p ).
\end{eqnarray}

\noindent
This integral is finite and non-singular at $p\to 0$, 
and, therefore, we have only a finite wave function
renormalization for  the $\pi_i$ fields. 

The correction to the mass of the pion superfield, $S_m$, is a sum of $D^2$-independent parts of the $I_1(p),I_2(p),I_3(p),I_{ct}$. It is given by the  relation

\begin{eqnarray}
S_m&=
\int\frac{d^3k}{(2\pi)^3}\Big\{
\frac{1}{k^2}(2-N)(1-f)+\frac{1}{k^2-m^2}(f-2)+
\frac{e^{2ik \wedge p}}{k^2}(N-2)(1-f)+
\frac{e^{2ik \wedge p}}{k^2-m^2}(2-f)
\Big\}
\end{eqnarray}

\noindent
and is both UV finite and without dangerous IV/UV mixing if
$(2-N)(1-f)=2-f$ i. e., if $f=N/(N-1)$. At this value of $f$, we
get  

\begin{equation}
S_m= m^2\frac{2-N}{N-1}\int\frac{d^3k}{(2\pi)^3}\frac{1-\cos(2k\wedge p)}
{k^2(k^2-m^2)},
\end{equation}

\noindent
which vanishes at $p=0$, as required.

The above result suggest that, since $f\to 1$ for large $N$,  only
the first quartic ordering  in (\ref{n1}) survives in this limit. This
indicates that 
the use of the $1/N$ expansion  becomes appropriated and that
proceeding along the lines described  in \cite{sigma}   one may
prove renormalizability to all orders.

Our analysis can be easily extended to the case in which the basic
superfields belong to a representation of the $U(N)$
group. For the fundamental representation, similarly  to the nonsupersymmetric
situation, one finds that Goldstone's theorem holds if a
gauge invariant Moyal ordering of the basic fields is adopted. For the
adjoint representation there are additional difficulties due to the matrix
character of the superfields.

\section{Generic Chiral Model in $3+1$ Dimensions}

As discussed in Section I, we shall next consider a generic $O(N)$ chiral
model whose action is given by

\begin{eqnarray}
\label{4act}
S=\int d^8 z \bar{\Phi}_a\Phi_a-\Big(\int d^6 z W(\Phi_a)+h.c.\Big), 
\end{eqnarray}

\noindent
where $W$ is the superpotential. In order to be invariant under local 
transformations we must have an even number of superfields, so that we can
take complex combinations of them. As for the Moyal ordering it must
be of the form  
$\dots * \Phi_a * \Phi_a * \Phi_{a+1} * \Phi_{a+1}* \dots$. 
We, then, restrict $N$ to be even and the superpotential to be given by 

\begin{eqnarray}
\label{4pot}
W(\Phi_a)=(-\frac{\mu}{2})\Phi_a  \Phi_a+
\sum_{k=2}^{\infty}\frac{\lambda_k}{k}(\Phi_a * \Phi_a)^{*k}.
\end{eqnarray}

 The consideration of other orderings brings no essential
modifications since, as we will show,  the vanishing of the two
point function of the pion field at zero momentum is completely
enforced by the chirality of the pion superfield.

\noindent
For a constant chiral superfield, satisfying the condition $\Phi_a \Phi_a = a^2$, the 
minimum of the superpotential is achieved for 

\begin{eqnarray}
\label{cond}
-\frac{\mu}{2}+\sum_{k=2}^{\infty}\lambda_k{(a^2)}^{k-1}=0.
\end{eqnarray}

\noindent
As usual, to break the $O(N)$ symmetry we perform the shift
$\Phi_a \rightarrow (\pi_i,\sigma+a)$ after which the superpotential can be
rewritten as

\begin{eqnarray}
\label{Wcorrec}
&&W(\pi_i,\sigma)= -\frac{m}{2}\sigma^2 + \sum_{k=2}^{\infty}\frac{\lambda_k}{k}\Big(
\big[\frac{(2k)!}{3!(2k-3)!} a^{2k-3}\sigma*\sigma*\sigma+
\frac{(2k)!}{4!(2k-4)!} a^{2k-4}\sigma*\sigma*\sigma*\sigma\big]\nonumber\\&+&
ka^{2k-4}\big[2(k-1)a\sigma+\frac{(2k-2)(2k-3)}{2}\sigma*\sigma\big]* \pi_i * \pi_i
+ 
\frac{k(k-1)}{2}a^{2(k-2)}(\pi_i*\pi_i)^2\Big)+\ldots,
\end{eqnarray}

\noindent
where only quadratic, cubic and quartic terms have been explicitly displayed. 
Then the field $\sigma$ acquires a mass

\begin{eqnarray}
\label{em}
m=-\mu + \sum_{k=2}^{\infty}\frac{\lambda_k}{k} \frac{(2k)!}{(2k-2)!} a^{2k-2},
\end{eqnarray}
while the kinetic term takes the form $\int d^8 z
(\bar{\pi}_i\pi_i+\bar{\sigma}\sigma)$. 

To cancel the divergences we introduce a counterterm action of the
form 

\begin{eqnarray}
S_{ct}&=&\int d^8 z \delta_z(\bar{\pi}_i\pi_i+\bar{\sigma}\sigma)+\Big\{\int d^6 z\Big[
(-\frac{\delta_\mu}{2}+
\sum_{k=2}^{\infty}\frac{\delta\lambda_k}{k} \frac{(2k)!}{2(2k-2)!}
a^{2k-2})\sigma^2+\nonumber\\&+& 
(-\delta_\mu a+2\sum_{k=2}^{\infty}\delta\lambda_k a^{2k-1})\sigma+
(-\frac{1}{2}\delta_\mu a+\sum_{k=2}^{\infty}\delta\lambda_k
a^{2k-1})\pi_i\pi_i+\nonumber\\&+&
\sum_{k=2}^{\infty}\frac{\delta\lambda_k}{k}\Big(
\big[\frac{(2k)!}{3!(2k-3)!} a^{2k-3}\sigma*\sigma*\sigma + \frac{(2k)!}{4!(2k-4)!}
a^{2k-4}\sigma*\sigma*\sigma*\sigma+\ldots\big]+ \nonumber\\
&+&k\big[2(k-1)a^{2k-3}\sigma+\frac{(2k-2)(2k-3)}{2}a^{2k-4}\sigma*\sigma\big]*\pi_i*\pi_i+
\nonumber\\&+&
\frac{k(k-1)}{2}a^{2(k-2)}(\pi_i*\pi_i)^2\Big)+\ldots
\Big]+h.c.\Big\}.
\end{eqnarray}

\noindent
The nonvanishing propagators for the chiral superfields $\pi_i$ and
$\sigma$ read, as usual,
 
\begin{eqnarray}
&&<\pi_i\bar{\pi}_j>=\delta_{ij}\frac{1}{\Box}\delta^8(z_1-z_2),\,\nonumber\\
&&<\sigma\bar{\sigma}>=\frac{1}{\Box-m^2}\delta^8(z_1-z_2),\qquad
<\sigma\sigma>=\frac{mD^2}{4\Box(\Box-m^2)}\delta^8(z_1-z_2).
\end{eqnarray}

\noindent
There are also $D^2,\bar{D^2}$ factors associated with the chiral 
vertices according to the standard Feynman rules in superspace
\cite{BK0}. Also observe that the propagator $<\pi_i\pi_j>=0$. 

The $<\pi_i\bar{\pi}_j>$ propagator will be represented by a dashed
line and the $<\sigma\sigma>$ propagator by a solid one. 
The $\bullet$ ($\circ$) symbol corresponds to a factor $D^2$ ($\bar{D}^2$) associated
with $<\bar{\sigma}\bar{\sigma}>$ ($<\sigma\sigma>$) propagator while all other
$D$-factors are associated with vertices by the usual rules in
superspace \cite{BK0}. The trigonometric factors in the vertices have the same structure as
in the models seen before. The vertex $\sigma*\pi_i*\pi_j$ carries a factor 

\begin{eqnarray}
\delta_{ij}\cos(p_1\wedge p_2),
\end{eqnarray}

\noindent
while for the vertex $\sigma*\sigma*\sigma$ one finds

\begin{eqnarray}
(\cos(p_1\wedge p_2)+\cos(p_2\wedge p_3)+\cos(p_1\wedge p_3)),
\end{eqnarray}

\noindent
where $p_1,p_2,p_3$ are three incoming momenta. The vertex
$\pi_i*\pi_j*\pi_k*\pi_l$ has a factor

\begin{eqnarray}
&&-\lambda[ \delta_{ij}\delta_{kl}(\cos(p_1\wedge p_2)\cos(p_3\wedge p_4)
+ \delta_{ik}\delta_{jl}(\cos(p_1\wedge p_3)\cos(p_2\wedge p_4)
\nonumber\\
&&+\delta_{il}\delta_{jk}(\cos(p_1\wedge p_4)\cos(p_2\wedge p_3))
], 
\end{eqnarray}

\noindent
whereas for the vertex $\pi_i*\pi_j*\sigma*\sigma$ {it turns out to be} 

\begin{eqnarray}
\delta_{ij}\cos(p_1\wedge p_2)\cos(p_3\wedge p_4).
\end{eqnarray}

\noindent
Furthermore, the $\pi$ field mass counterterm and the $\sigma$ tadpole
counterterm are, respectively, 

\begin{eqnarray}
&&-\delta_{ij}(-\delta_{\mu}a
  +2\sum_{k=2}^{\infty}\delta\lambda_ka^{2k-1}) \qquad \mbox{and}\,\,\,\,\,\,\,\,\,\,
\delta_{\mu}a-2\sum_{k=2}^{\infty}\delta\lambda_k a^{2k-1}.
\end{eqnarray}

\noindent
The gap equation signalizing that $\sigma$ has vanishing expectation
value is depicted in Fig. 4. 

\noindent Notice that the supergraph including one $\sigma\pi_i\pi_j$ vertex
vanishes because $<\pi_i(z_1)\pi_j(z_2)>=0$. 
The gap equation is then the same as in the commutative case. 
The tadpole graph gives a contribution proportional to

\begin{eqnarray}
m\int d^6 z \sigma(z)\frac{\bar{D}^2D^2\bar{D}^2}{64\Box(\Box-m^2)}
\delta^8(z-z')|_{z=z'}=
-m\int d^8 z \sigma(z)\frac{1}{(\Box-m^2)}\delta^4(x-x')|_{x=x'}=0, 
\end{eqnarray}

\noindent
which vanishes like in the commutative case. The vanishing is only due
to supersymmetry, since all noncommutative trigometric factors have 
disappeared. Therefore, vanishing of the couterterm contribution leads
to 

\begin{equation}
\delta_{\mu}a-2\sum_{k=2}^{\infty}\delta\lambda_ka^{2k-1}=0,
\end{equation}

\noindent
fixing a relation among the counterterms. It also coincides with
the pion mass counterterm implying that it vanishes. 

The only one-loop contribution to the pion mass renormalization is
given by the supergraph shown in Fig. 5.
The contribution of this supergraph turns out to be
proportional to 

\begin{eqnarray}
\int d^8 z \pi_i(z)\pi_i(z)\frac{1}{(\Box-m^2)}\delta^8(z-z')|_{z=z'},
\end{eqnarray}

\noindent
which also vanishes due to the chirality of $\pi_i$.  Had we used
other orderings this expression would contain non trivial phase factors
but, nevertheless, still would vanish by the same reason (i.e., the chirality
of the pion field).
Therefore we conclude that at one-loop order there is no mass
renormalization and this is entirely due to the chirality of the
superfield $\pi_i$. This is in contrast to the supersymmetric case in
three dimensions where a nonvanishing contribution was found. There the
superfields are real and no chirality argument is available. 

Just using supersymmetry and chirality arguments allow us to go beyond one
loop straightfowardly. At two loops we have to
consider supergraphs with two $\pi_i$ external lines
which are drawn in Fig. 6. The supergraphs in Figs.
6$a$, 6$b$, 6$c$ and 6$d$ contain  tadpole-like loop contributions
consisting of one propagator $<\sigma(z)\sigma(z)>$ or
$<\bar{\sigma}(z)\bar{\sigma}(z)>$. Since they are proportional to
$D^2\bar{D}^2D^2\delta^8(z-z')|_{z=z'}$ or  
$\bar{D}^2D^2\bar{D^2}\delta^8(z-z')|_{z=z'}$, they both vanish. 

The supergraphs in Figs. 6$e$, 6$f$, 6$g$, 6$h$, 6$i$, 6$j$ and 6$k$
have equal numbers of $D^2$ and
$\bar{D}^2$ factors, hence, after $D$-algebra transformations, their
contributions are proportional to $\int d^8 z \pi_i\pi_i$ which is equal to
zero due to chirality of $\pi_i$.
The supergraphs depicted in Figs. 6$l$, 6$m$ and 6$n$ are proportional to
$D^2\delta_{12}D^2\delta_{12}$ and to 
$\bar{D}^2\delta_{12}\bar{D}^2\delta_{12}$. Both structures give vanishing
contributions after integration by parts. 
Finally, the supergraph drawn in Fig. 6$o$
is proportional to $\int d^4\theta \pi_i D^2\pi_i=\int
d^2\theta\pi_i\Box\pi_i$ and vanishes for slowly varying superfields.
We then conclude that there is no loop corrections to the pion mass
for slowing varying superfields. Then the Goldstone theorem is
satisfied at one and two loops order for the $O(N)$ supersymmetric chiral
superfield whose interactions are compatible with gauge
invariance. Notice that the vanishing of these corrections is due to
the supersymmetry and not to relations among the coupling
constants. In fact, this is a generalization of the nonrenormalization
theorems of supersymmetric theories \cite{BK0} to the noncommutative
context. 

We can also consider the purely noncommutative interaction
(\ref{epsilon}) in superspace. Its contribution to (\ref{Wcorrec})
is similar to (\ref{eps}) but its loop constributions always vanish
since $<\pi_i \pi_j>=0$. So it is also not relevant for the results
just derived.

\section{Conclusions}

In this work, the mechanism of spontaneous symmetry breaking and the
appearance of Goldstone bosons were investigated in connection with
several noncommutative field models.  

We first studied the spontaneous breakdown of the continuous $O(N)$
symmetry in a noncommutative scalar model with quartic and sextuple
interactions. For $2+1$-dimensions, there is a class of fields
ordering for which the model turned out to be renormalizable, up to
one loop. The linear combination of couterterms fixed by the gap
equation equals the $\pi$-field mass counterterm. Moreover, for the
pion two point function, a single condition suffices to ensure: i)
cancellation of the UV divergences, ii) cancellation of the IR
divergences arising from the UV/IR mixing, iii) appearance of massless
excitations. As shown, this is a peculiarity of the $2+1$-dimensional
world. 
In $3+1$-dimensions renormalizability called for the elimination of the
sextuple interactions while the $UV$ convergence of the planar part
restricts the dynamics as already found in \cite{KK}. In
$2+1$-dimensions supersymmetry restrics even further the class of
allowed models while, on the other hand, renders the theory
renormalizable to all orders in the limit $N \rightarrow \infty$. 

Secondly, we studied a generic chiral model in $3+1$-dimensions with a
local $O(N)$ gauge theory. We argue that this model, although being
not renormalizable by power counting, may provide a realistic
description of the low energy limit of compactified string or
M-theory. There are no one loop corrections to the pion mass while the
same holds for two loops in the limit of slowly varying
superfields. Supersymmetry enabled us to go through the $N = 2$
barrier existing for purely bosonic interactions\cite{KK}. 

We also found a purely noncommutative interaction which preserves
the $O(N)$ symmetry and is present in any dimension. It starts
contributing at $N-2$ loops so that it is not relevant for our one
loop analysis. It clearly deserves further study for the understanding
of its properties.

Also relevant is the phenomenon of spontaneous breaking of gauge
symmetry in noncommutative supersymmetric gauge theories. Its study is
in progress. 

\section{Acknowledgments}

This work was partially supported by Funda\c c\~ao de Amparo \`a
Pesquisa do Estado de S\~ao Paulo (FAPESP) and Conselho Nacional de
Desenvolvimento Cient\'\i fico e Tecnol\'ogico (CNPq). H. O. G. and
V. O. R. also acknowledge support from PRONEX under contract CNPq
66.2002/1998-99. A.Yu. P. has been supported by FAPESP, project No. 00/12671-7.

\begin{figure}[htbp]
\includegraphics{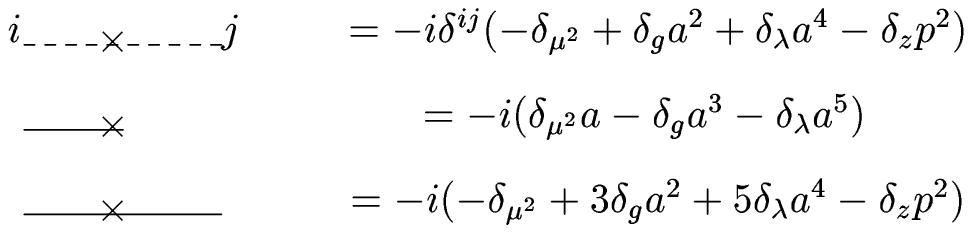}
\caption{Structure of counterterms in the noncommutative $\phi^6$
model. Continuous and dashed lines represent the $\sigma$
and $\pi_i$ fields respectively.}  
\end{figure}
\vspace*{1.5cm}
\begin{figure}[htbp]
\includegraphics{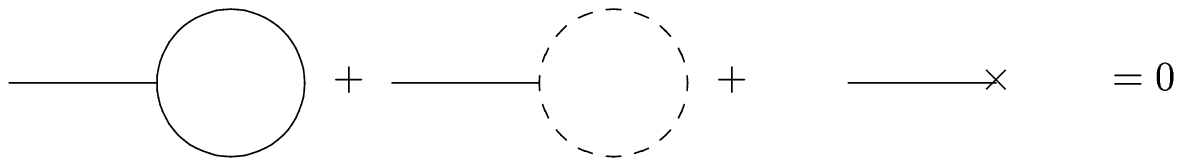}
\caption{Gap equation in the noncommutative $\phi^6$ model.}
\end{figure}
\vspace*{2cm}
\begin{figure}[htbp]
\includegraphics{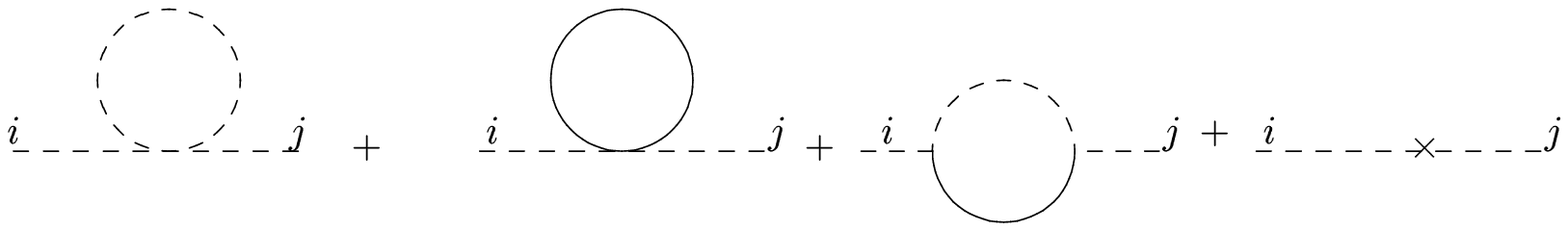}
\vspace*{6mm}
\caption{One-loop contributions to the pion's two-point function in the
noncommutative $\phi^6$ model.}
\vspace*{0.5cm}
\includegraphics{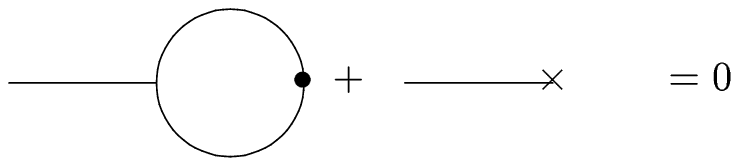}
\vspace*{5mm}
\caption{Gap equation in generic chiral model in $3+1$ dimensions.}
\includegraphics{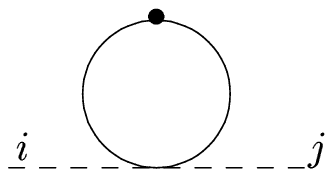}
\vspace*{4mm}
\caption{One-loop contribution to the pion's two-point function 
in generic chiral model in $3+1$ dimensions.}
\end{figure}


\begin{figure}[ht]
\includegraphics{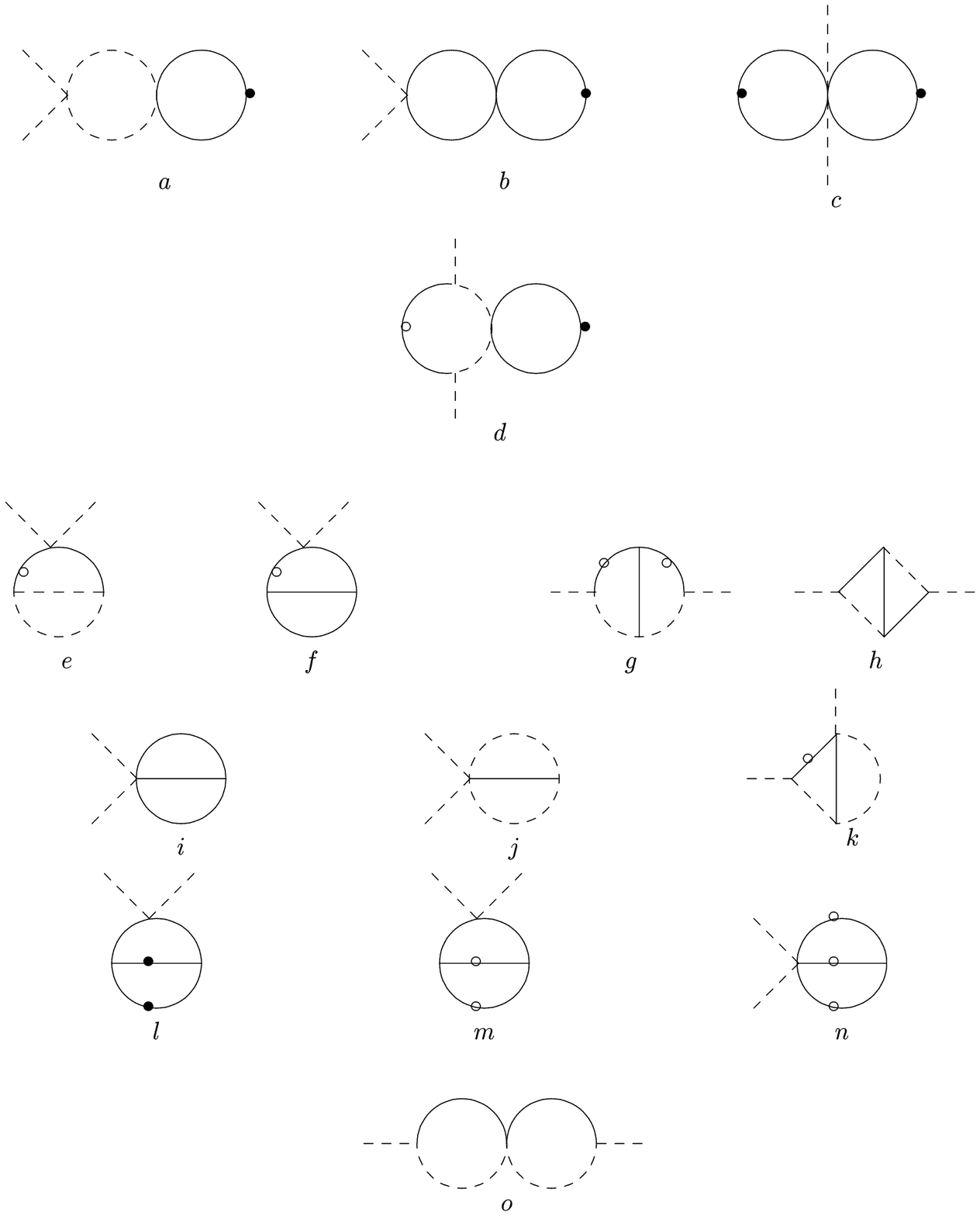}
\vspace*{4mm}
\caption{Two-loop contributions to the pion's two-point function 
in generic chiral model in $3+1$ dimensions.}
\end{figure}
\end{document}